\documentclass[structabstract]{aa} 
\usepackage{graphicx}
\usepackage{txfonts}
\usepackage{natbib}
\usepackage{longtable}
\usepackage{lscape}
\usepackage{multirow}
\usepackage{color}
\usepackage{lscape}

\newcommand{\kms}{km\,s$^{-1}$}

\begin{document}

\title{Large-field CO\,(1--0) observations toward the Galactic historical supernova remnants: a large cavity around Tycho's supernova remnant}
\author{X.~Chen\inst{1}, F.~Xiong\inst{1,2}, and J.~Yang\inst{1}}
\institute{Purple Mountain Observatory \& Key Laboratory of Radio Astronomy, Chinese Academy of Sciences, 2 West Beijing Road, 210008 Nanjing, PR China\\
e-mail: xpchen@pmo.ac.cn
\and 
University of Chinese Academy of Sciences, No. 19A Yuquan Road, 100049 Beijing, PR China}

\date{}

\abstract{The investigation of the interaction between the supernova remnants (SNRs) and interstellar gas is not only necessary 
to improve our knowledge of SNRs, but also to understand the nature of the progenitor systems.}
{As a part of the Milky Way Imaging Scroll Painting CO line survey, the aim is to study the interstellar gas surrounding the 
Galactic historical SNRs. In this work, we present the CO results of Tycho's SNR.}
{Using the 3$\times$3 Superconducting Spectroscopic Array Receiver (SSAR) at the PMO 13.7-meter telescope, we performed
large-field (3$^\circ$\,$\times$\,2$^\circ$) and high-sensitivity CO\,(1--0) molecular line observations toward Tycho's SNR.}
{The CO observations reveal large molecular clouds, stream-like structures, and an inner rim around the remnant. We derived the 
basic properties (column density, mass, and kinematics) of these objects based on the CO observations. The large molecular 
clouds individually show an arc toward the remnant center, outlining a large cavity with radii of $\sim$\,0.3$^\circ$\,$\times$\,0.6$^\circ$ 
(or 13\,pc\,$\times$\,27\,pc at a distance of 2.5\,kpc) around the remnant. The CO line broadenings and asymmetries detected 
in the surrounding clouds, the observed expansion of the cavity, in concert with enhanced $^{12}$CO\,(2--1)/(1--0) intensity ratio 
detected in previous studies, suggest the interaction of the large cavity with a wind in the region. After excluding the scenario of 
a large bubble produced by bright massive stars, we suggest that the large cavity could be explained by accretion wind from 
the progenitor system of Tycho's supernova. Nevertheless, the possibility of the random distribution of a large cavity around 
Tycho's SNR cannot be ruled out thus far. Further observations are needed to confirm the physical association of the large 
cavity with Tycho's SNR.}
{}

\authorrunning{Chen et al.}
\titlerunning{CO observations toward Tycho's SNR}

\maketitle

\section{Introduction}

Type Ia supernovae (SNe~Ia) are thermonuclear explosions of white dwarfs (WDs) in binary systems. Nevertheless, the specific 
progenitor systems of the SNe Ia have not been identified (Hillebrandt \& Niemeyer 2000; Maoz et al. 2014). Two broad classes 
of progenitor binary systems have been suggested: the single-degenerate (SD) scenario (Whelan \& Iben 1973; Nomoto 1982), 
in which the WD accretes mass from a non-degenerate stellar companion (e.g., main sequence or red giant star) and explodes 
when it exceeds the Chandrasekhar mass; and the double-degenerate (DD) scenario (Iben \& Tutukov 1984; Webbink 1984), 
involving the merger of two WDs. 

The supernova of 1572 (SN~1572), also widely known as ``Tycho's supernova", is a well-established Type Ia supernova, 
verified from its light echoes (Rest et al. 2008; Krause et al. 2008). As one of the few historical SNRs in the Milky Way, Tycho's 
SNR has been widely observed over the entire electromagnetic spectrum (see, e.g., Warren et al. 2005; Gomez et al. 2012). 
These observations have found in the remnant a complete shell-like structure with a diameter of approximately 8\,arcmin, 
produced by the shocks from the SN explosion (see Warren et al. 2005). The distance of Tycho's SNR has been estimated 
to be between 2 and 5\,kpc, but recent studies suggest a value closer to 2.5 and 3\,kpc (Tian \& Leahy 2011; Zhang et al. 2013).

Multi-wavelength observations suggested the existence of circumstellar material (CSM), including gas (Reynoso et al. 1999; 
Lee et al. 2004; Cai et al. 2009; Xu et al. 2011) and dust (Ishihara et al. 2010; Gomez et al. 2012), in the northeast and east 
front of Tycho's SNR shell. Nevertheless,  Tian \& Leahy (2011) suggest that the atomic hydrogen (HI) gas at velocities 
of $-$47 to $-$53\,km\,s$^{-1}$, which was once considered to be interacting with the shock waves from the explosion (Reynoso 
et al. 1999), is located in front of Tycho's SNR. These authors also argued that the CO molecular gas at a velocity 
of $\sim$\,$-$\,64\,km\,s$^{-1}$ (Lee et al. 2004; Cai et al. 2009; Xu et al. 2011) is not interacting with the remnant either, 
because the derived gas density from the CO observations ($\sim$\,200\,cm$^{-3}$) is much larger than the density referred 
from the X-ray observations ($\sim$\,0.2\,cm$^{-3}$; Katsuda et al. 2010). However, the re-analysis of the high-energy observations 
(including X- and $\gamma$-ray data) of Tycho's SNR suggested a denser ambient medium ($\sim$\,4-12 cm$^{-3}$ 
on average; Zhang et al. 2013). 
Recently, Zhou et al. (2016) found that the CO gas around the shell is expanding and that there is an enhanced $^{12}$CO\,(2--1)/(1--0) 
intensity ratio ($\sim$\,1.6) in the northeast front of the shell (hereafter we refer to the circumstellar CO gas located at the edges
of the 8$'$ diameter shell as the rim). These new observational results provide evidence for the interaction between the supernova 
shocks and the molecular gas near the remnant. In addition, this interaction is believed to be responsible for the acceleration of 
cosmic-ray protons detected in Tycho's SNR (see, e.g., Zhang et al. 2013).

It should be noted that most of the previous observations toward Tycho's SNR focused on  only a limited region (less than 30 
arcmin) around the shell-like structure. In this paper, we present large-field (3$^\circ$\,$\times$\,2$^\circ$) and high-sensitivity 
CO\,(1--0) molecular line observations toward Tycho's SNR, using the 13.7-meter millimeter-wavelength telescope of the 
Purple Mountain Observatory (PMO), which is part of the Milky Way Imaging Scroll Painting (MWISP) project for investigating 
the nature of the molecular gas along the northern Galactic Plane. In Section 2 we describe the observations and data reduction. 
Observational results are presented in Section 3, discussed in Section 4, and summarized in Section 5.

\section{Observations and data reduction}

\subsection{PMO 13.7-meter CO observations}

The CO\,(1--0) observations toward Tycho's SNR is part of the MWISP CO line survey project operated by the PMO, which were made 
from 2011 November to 2016 February with the 13.7-meter millimeter-wavelength telescope of the Qinghai station of PMO at Delingha in 
China. The nine-beam Superconducting Spectroscopic Array Receiver (SSAR) was working as the front end in sideband separation mode 
(see Shan et al. 2012). Three CO\,($J$ = 1--0) lines were simultaneously observed, $^{12}$CO at the upper sideband (USB) and two other 
lines, $^{13}$CO and C$^{18}$O, at the lower sideband (LSB). Typical system temperatures were around 210\,K for the USB and around 
130\,K for the LSB, and the variations among different beams are less than 15\%. 
The total of pointing and tracking errors is about 5$''$, while the half-power beam width (HPBW) is $\sim$\,52$''$. The main-beam efficiencies 
during the observations were $\sim$\,44\% for USB with the differences among the beams less than 16\%, and $\sim$\,48\% for LSB with the 
differences less than 6\%. We mapped a 3$^\circ$\,$\times$\,2$^\circ$ area around Tycho's SNR via on-the-fly (OTF) observing mode, and 
the data were meshed with the grid spacing of 30$''$. A Fast Fourier Transform (FFT) spectrometer with a total bandwidth of 1000\,MHz and 
16,384 channels was used as the back end. The corresponding velocity resolutions were $\sim$\,0.16\,km\,s$^{-1}$ for the $^{12}$CO line 
and $\sim$\,0.17\,km\,s$^{-1}$ for both the $^{13}$CO and C$^{18}$O lines. The average rms noises of all final spectra are about 0.5\,K for 
$^{12}$CO and about 0.3\,K for $^{13}$CO and C$^{18}$O. All data were reduced using the GILDAS package (see 
https://www.iram.fr/IRAMFR/GILDAS/).

\subsection{Complementary CO data}

A number of large CO line surveys along the Galactic Plane have been carried out in the past (see a review by Heyer \& Dame 2015), such 
as the $^{12}$CO\,(1--0) surveys at the Center-for-Astrophysics (CfA) 1.2-meter telescope (Dame et al. 2001) and the Five College Radio 
Astronomy Observatory (FCRAO) 14-meter telescope (Heyer et al. 1998). These previous surveys also obtained large-field $^{12}$CO images 
around Tycho's SNR, which provided complementary CO data in this work.

\section{Results}

Figure~1 shows the velocity-integrated intensity image of the MWISP $^{12}$CO\,(1--0) emission around Tycho's SNR. 
The integrated velocity range is between $-$68 and $-$57\,km\,s$^{-1}$, a velocity range suggested by previous CO line 
observations toward the shell-like structure in the remnant (Lee et al. 2004; Cai et al. 2009; Xu et al. 2011; Zhou et al. 2016). 
The large-field MWISP $^{12}$CO image reveals three large molecular clouds, located in the southeast, northeast, and west 
of the remnant, respectively. Interestingly, the three clouds individually show an arc toward the remnant center (see solid 
yellow lines in Fig.\,1). The three arcs could be fitted by one complete ellipse, which outlines a previously undiscovered cavity 
around Tycho's SNR (see Fig.\,1). The radii of the cavity, depending on the azimuths, range from $\sim$\,0.3 degrees (to 
the south of the remnant) to $\sim$\,0.6 degrees (to the west of the remnant), corresponding to $\sim$\,13-27\,pc (adopting a 
distance of 2.5\,kpc).

The MWISP $^{12}$CO image also reveals three stream-like structures in the cavity (see Fig.\,1). One is to the northeast of the 
remnant (named stream$_{\rm NE}$), the other, showing a wiggle morphology, to the southeast (named stream$_{\rm SE}$). 
These two stream-like structures spatially connect the east edge of the cavity and the inner rim, which was found in the 
previous CO line observations (with small fields of view) toward Tycho's SNR (e.g., Lee et al. 2004; Zhou et al. 2016).
Another faint stream-like structure can be roughly distinguished to the northwest of the remnant (named stream$_{\rm NW}$). 
The three ``streams", joining at the center of the remnant, appear to radiate from the center.

Figure~2 shows the velocity-integrated intensity image of the MWISP $^{13}$CO\,(1--0) emission around Tycho's SNR. The 
$^{13}$CO emission also shows the three clouds, but the emission is faint and the arc-like structures seen in the $^{12}$CO
emission are not clear in the $^{13}$CO intensity image. The $^{13}$CO emission is also detected from the north rim around the 
shell-like structure. Figure~3 shows the CO spectra sampled from six positions in the surrounding clouds (see Figure~2). The 
CO spectra do not show the Gaussian-like shape which is generally seen in quiescent molecular clouds, but present line 
broadenings (in a velocity range between $-$70 and $-$55\,\kms) and asymmetries (see Figure~3). We note that no C$^{18}$O 
line emission is detected around Tycho's SNR in the MWISP observations. The MWISP $^{12}$CO velocity channel map is 
shown in Fig.\,A.1, where detailed kinematic information of the molecular gas around Tycho's SNR can be found. 

The physical properties of the clouds, stream-like structures, and inner rim are listed in Table~1. Here two methods have been 
used in the derivation of the H$_{2}$ gas column densities and masses. In the first method, on the assumption of local thermodynamic 
equilibrium (LTE) and the $^{12}$CO\,(1--0) line being optically thick, we can derive the excitation temperature from the peak 
radiation temperature of the $^{12}$CO\,(1--0). The $^{13}$CO\,(1--0) emission is optically thin and the $^{13}$CO column density 
is converted to the H$_2$ column density using N(H$_2$)/N($^{13}$CO) $\approx$ 7\,$\times$\,10$^5$ (Frerking et al. 1982). 
In the second method, the H$_2$ column density is estimated by adopting the mean CO-to-H$_2$ mass conversion factor 
1.8\,$\times$\,10$^{20}$\,cm$^{-2}$\,K$^{-1}$\,km$^{-1}$\,s (Dame et al. 2001). The difference between the H$_2$ column 
densities derived by the two methods are mainly caused by the small filling factors of the $^{13}$CO emission, since there is much 
less $^{13}$CO emission than $^{12}$CO emission in the region (see Figures~1 and 2). 

\section{Discussion}

\subsection{Scenario of a cavity produced by bright massive stars}

The large-field CO\,(1--0) observations show a large cavity around Tycho's SNR. It is well-known that bright (O- or early B-type) 
stars are able to produce large bubbles in the ISM through strong stellar winds and UV radiation (see, e.g., Churchwell et al. 2006; 
Deharveng et al. 2010). Therefore, the first question is whether the cavity around Tycho's SNR was produced by the bright stars 
in the region.

As found by Chen et al. (2013), there is a linear relationship between the radius of a main-sequence bubble in a molecular environment 
($R_{\rm bubble}$) and the initial mass of the energy source star ($M_{\rm star}$): 
$R_{\rm bubble}$ (pc) $\approx$ 1.22$M_{\rm star}$/$M_\odot$ $-$ 9.16\,pc, assuming a constant interclump pressure (see Chen 
et al. 2013 for more details). For the large cavity found in the MWISP observations (major radius of $\sim$\,0.6$^\circ$ or 27\,pc), a 
massive star with a mass of $\sim$\,30\,$M_\odot$ (O7 or earlier types) is required. However, no such massive early stars are found 
in the center of the cavity in the SIMBAD Astronomical Database\footnote{http://simbad.u-strasbg.fr/simbad/}. In addition, there is no 
large HII region around the remnant either. Therefore, we can exclude the scenario of massive star bubble in the case of the large 
cavity around Tycho's SNR.

\subsection{Possibility of the random distribution of a cavity}

The progenitor of Tycho's supernova was an evolved WD system (the time delays of most SNe~Ia range from $\sim$\,275\,Myr
to 1.25\,Gyr, with a median of $\sim$\,650\,Myr; see Schawinski 2009), and would not be expected to remain associated with its 
natal molecular cloud. Therefore, the second question is whether the large cavity is randomly distributed around the remnant. 

After checking much larger CO images from previous surveys (e.g., the FCRAO data), we find that the cavity-like structure is commonly 
seen in the field and the estimated probability for a `cavity' to be detected by chance is higher than 0.06 (see Appendix~B). On the other 
hand, we note that line broadenings and asymmetries are detected in the CO spectra of the surrounding clouds (see Figure 3). For instance, 
the measured $^{12}$CO FWHM linewidths in the southeast cloud are typically $\sim$\,6-7\,\kms, which are much broader than the 
linewidths found in other molecular clouds, such as the L1157 cloud in Taurus ($\sim$\,0.5-1.0\,\kms; see Hacar et al. 2016), for example. 
This kind of spectra, suggesting the shock effect on the surrounding gas, is frequently found in the molecular clouds interacting with SNRs 
(see, e.g., Jiang et al. 2010, Zhou et al. 2014). Furthermore, the observed kinematics in the CO gas suggests that the whole cavity is 
expanding at a velocity of $\sim$\,3-4\,\kms\ (see discussion below in $\S$\,4.4.1). These results, in concert with the enhanced 
$^{12}$CO(2--1)/(1--0) line ratio (about 1.6) found in the rim in the previous studies (Zhou et al. 2016; see discussion below in $\S$\,4.4.3), 
suggest the interaction of the large cavity with a strong wind in the region.

Nevertheless, it must be noted that we cannot rule out the possibility of the random distribution of a large cavity-like structure around 
 Tycho's SNR thus far. Further observations, for example, searching for 1720\,MHz OH maser around the remnant (see discussion in 
 Chen et al. 2014 and Dubner \& Giacani 2015), are needed to verify the physical association between the large cavity and remnant.

\subsection{Potential cavity opened by accretion wind} 

According to theoretical studies (e.g., Hachisu et al. 1996; 1999), accreting WDs in a binary system with a non-degenerate 
companion would blow substantial outflows (also known as ``accretion winds''), excavating low-density cavities in the surrounding 
interstellar medium (ISM) in the few 10$^6$\,yr prior to explosion (see Badenes et al. 2007 and references therein).  On the other 
hand, in the DD scenario, there would be no such winds and cavities. Therefore, one direct and effective way to distinguish between 
progenitor scenarios is to search for large cavities (10-30\,pc) in the ISM, centered on the explosion sites of the SNe Ia. However, 
to our knowledge, no such cavities have been observationally found yet (see, e.g., Badenes et al. 2007). A promising candidate 
could be RCW~86, an SNR located in a 12-pc-radius cavity (see Broersen et al. 2014 and references therein), though the Type Ia 
origin of this remnant is still uncertain. 

The dimension of the large cavity around Tycho's SNR is far larger than the shock front from the supernova could produce. 
In this context, this large cavity could be explained by the accretion wind from the progenitor system of Tycho's supernova. 
As mentioned above, the progenitor system of Tycho's supernova was an evolved system. The molecular clouds found 
around  Tycho's progenitor system are likely newly formed in the Galactic Plane, where the typical lifetime of molecular 
clouds is $\sim$\,10-100\,Myr (see Heyer \& Dame 2015). These newly-formed clouds, affected by generations of supernovae, 
stellar winds, colliding flows, and turbulence (see, e.g., Dobbs et al. 2014), are irregularly distributed around Tycho's progenitor 
system and then impacted by its accretion wind. This may explain why the north and northwest part of the cavity is open, that is, 
there are no molecular clouds initially formed/distributed in these directions. Interestingly, a similar situation is also seen in the RCW\,86 
case, where CO clouds are only seen in the east, south, and northwest of the cavity, leaving most of the circumference open 
(see Sano et al. 2016).

Nevertheless, the CO observations (see Figs.\,1 and 2) find inner rim and stream-like structures within the cavity. This raises 
another question about how the progenitor system can excavate a large cavity through a strong wind whilst leaving a significant 
amount of residual material in the center.
It should be noted that the cavity models discussed in Badenes et al. (2007) make three important approximations: the ISM 
around the progenitor system is homogeneous, and accretion wind from the system is spherically symmetric in space and 
continuous in time. As we discussed above, the interstellar gas around the progenitor systems could be very complicated. 
For accretion winds, in fact, numerical studies have shown that (1) the winds could be bipolar, which may lead to a bipolar 
CSM structure similar to planetary nebulae (Balick \& Frank 2002), and (2) the winds are normally not steady but instead 
episodic (e.g., Hachisu \& Kato 2003a, 2003b). Furthermore, during the mass growth of the WD, the mass transfer between 
the donor star and WD is unstable, where nova-like outbursts are expected (see Maoz et al. 2014 and references therein) 
and will lead to a complex CSM structure (even with several shells). Therefore, a totally empty cavity is only an ideal situation.

\subsection{Kinematics of the gas around the remnant} 

\subsubsection{The expansion of the surrounding clouds}

Figure~4 shows the $^{12}$CO intensity-weighted velocity field around Tycho's SNR. In this velocity-field image, the mean 
local standard of rest velocities ($V_{\rm LSR}$) of the molecular gas around the remnant can been found. The velocity field 
shows that the CO emission from the most part of the inner rim, as well as the stream-like structures, is relatively blueshifted (i.e., 
moving toward us), compared with the nearby clouds. The measured mean velocity of the inner rim is $-$65.0\,$\pm$\,0.5\,\kms, 
while the velocity measured along the cavity edge is roughly $-$62\,$\pm$\,1\,\kms\ (see Fig.\,4).

Figure~5 shows the position-velocity (PV) diagrams along the three large clouds (see the routes of the PV diagrams in Fig.\,4). 
Interestingly, the PV diagrams of the three clouds all show curve-shaped morphologies, which could be fitted by ellipses with 
velocity radii of $\sim$\,3--4\,\kms\ (see Fig.\,5). These results imply the expansion of the surrounding gas, which is likely pushed 
by the accretion wind. Furthermore, we find that the expansion directions of the northeast cloud (blueshifted) and west cloud 
(redshifted) appear to be opposite each other (see Fig.\,6). Therefore, the observed kinematics could by explained by a 
complete view that the large cavity is expanding.

Using the standard method (see Weaver et al. 1977), the value of the mechanical luminosity of the wind ($L_{\rm wind}$) can 
be estimated by 
$L_{\rm wind}$\,$\approx$\,$\frac{1}{3}$$\frac{n_{\rm gas}}{{\rm cm^{-3}}}$($\frac{R_{\rm c}}{{\rm pc}}$)$^2$($\frac{V_{\rm c}}{{\rm km\,s^{-1}}}$)$^3$\,$\times$\,
10$^{30}$~ergs\,s$^{-1}$, in order to excavate a cavity with a radius of $R_{\rm c}$ and expansion velocity of $V_{\rm c}$ in 
a molecular cloud with a density of $n_{\rm gas}$. For the large cavity around Tycho's SNR, the expansion velocity $V_{\rm c}$ 
is estimated to be $\sim$\,3.5\,km\,s$^{-1}$, while the density $n_{\rm gas}$ of the clouds is measured at $\sim$\,30\,cm$^{-3}$ 
from the CO observations. Adopting the major radius of the cavity ($R_{\rm c}$\,=\,27\,pc), the estimated 
$L_{\rm wind}$ ($\sim$\,3\,$\times$\,10$^{35}$\,ergs\,s$^{-1}$) could be fed by a wind with a mass-loss rate of the order of 
10$^{-6}$\,$M_{\odot}$\,yr$^{-1}$ at a velocity of $\sim$\,800\,\kms, while the timescale of the wind needed for opening such 
a large cavity ($\frac{16}{27}$$\frac{R_{\rm c}}{\rm pc}$$\frac{\rm km\,s^{-1}}{V_{\rm c}}$\,$\times$\,10$^6$\,yr; Weaver et al. 
1977) is $\sim$\,4-5\,$\times$\,10$^6$\,yr. The estimated wind velocity and timescale are both consistent with the predictions 
from the wind-regulated accretion models, in which the wind velocity ranges from $\sim$\,200\,\kms\ to $\sim$\,1000\,\kms\ 
and timescale is $\sim$\,10$^{6}$ yr (see Badenes et al. 2007 and references therein).

\subsubsection{The stream-like structures} 

The morphologies of the stream-like structures seen in the CO images resemble the jets/outflows driven by young stellar objects 
(YSOs; see, e.g., Reipurth \& Bally 2001; Arce et al. 2007). In the optical and infrared observations, those YSO jets can move 
more than 10\,pc from their driving sources, with a typical jet velocity of 100\,km\,s$^{-1}$ (see Reipurth \& Bally 2001). If the 
`streams' found in the cavity were indeed the YSO jets, there should be a small group of YSOs in the center of the remnant. However, 
we did not find this YSO group in the infrared observations (X.~Chen et al., in preparation).

In the wind-regulated accretion models (e.g., Hachisu et al. 1996; 1999), the accretion wind could last for a few million years, and 
the WD may explode as a Type Ia supernova while the accretion wind is still active. Therefore, one possible explanation is that 
these stream-like structures actually record the accretion winds from the progenitor system. In the theoretical models, the velocities 
of the accretion winds range from $\sim$\,200\,\kms\ up to $\sim$\,1000\,\kms\ (see discussion in $\S$\,4.4.1), which is high enough 
to shock interstellar clouds with large velocity dispersions. Indeed, in the CO observations (see PV diagrams in Figs.\,5a \& 5b), 
the stream-like structures show large velocity ranges ($\sim$\,10\,\kms). The observed result suggests shocked emission along the 
streams, which could be caused by the underlying wind. 

\subsubsection{The inner rim}

The  rim around the shell-like structure in Tycho's SNR was observed for decades (Lee et al. 2004; Cai et al. 2009; Xu 
et al. 2011), and was recently suggested to represent a wind-blown cavity ($\sim$\,8$'$ in diameter) from the progenitor 
system of the supernova, based on the detection of the expansion of the rim (see Zhou et al. 2016, and also Chiotellis et al. 
2013). The expansion velocity of the rim is $\sim$\,4.5\,km\,s$^{-1}$, while the estimated wind velocity and timescale for this 
putative cavity are $\sim$\,140\,km\,s$^{-1}$ and 3.9\,$\times$\,10$^5$\,yr, respectively (Zhou et al. 2016). 

In the MWISP CO intensity images (see Fig.\,7a), there is extended CO emission within the shell region, and a small cavity-like 
structure is found around the shell (as seen at the edges of the rim in Fig.\,7a, and in the green contours in Fig.\,2). Figure~7 
shows the PV diagrams across the rim but in various directions. The PV diagrams along the northeast rim (see Figs.\,5b \& 7c) 
show ring-like morphologies with small radii (about 0.1$^\circ$), which suggests the expansion of the gas in this direction with 
a velocity of $\sim$\,3.5--4.0\,\kms. This is consistent with the result found by Zhou et al. (2016). 

As discussed in $\S$\,4.3, during the accretion of the WD, nova-like outbursts could happen due to the instability in the
mass transfer between the donor star and WD. For instance, Langer et al. (2000) even found a long ($\sim$\,10$^6$\,yr) 
switch-on phase of the mass transfer in their binary models. Therefore, there is a possibility that the rim (found in previous 
observations) and the large cavity (found in this work) represent two independent cavities resulting from two different 
accretion/outburst epochs.

On the other hand, if the rim indeed represents a small cavity, we may expect to find the isotropic expansion of this small 
cavity. Nevertheless, we find that the expansion is only seen in the northeast direction of the rim, but not distinct in the other 
directions (see Figs.\,5a \& 7b). Alternatively, we suggest that the rim may trace a large amount of CSM swept-up by the 
(asymmetric) accretion wind in the northeast direction (see discussion above) for the following reasons: (1) The `expansion', 
as well as the most circumstellar gas, is seen in the northeast direction of the rim; (2) the rim is connected with the
stream-like structures (stream$_{\rm NE}$ and stream$_{\rm SE}$; see Fig.\,1) radiating from the remnant center toward the 
edges of the large cavity.

\section{Summary}

We present large-field CO\,(1--0) molecular line observations toward Tycho's SNR, using the PMO 13.7-meter telescope.
The CO images reveal, from the outside in, large molecular clouds, stream-like structures, and an inner rim around the remnant.
We derived the basic properties (column density, mass, and kinematics) of these objects based on the CO observations.

The large molecular clouds individually show an arc toward the remnant center, outlining a large cavity with radii of 
$\sim$\,0.3$^\circ$\,$\times$\,0.6$^\circ$ (or 13\,pc\,$\times$\,27\,pc at a distance of 2.5\,kpc) around the remnant. 
The observed CO line broadenings and asymmetries in the surrounding clouds, together with the enhanced $^{12}$CO\,(2--1)/(1--0) 
intensity ratio detected in previous observations, suggest the interaction of the large cavity with a wind in the region. 
After excluding the scenario of a large bubble produced by bright massive stars, we suggest that the large cavity could 
be explained by the accretion wind from the progenitor system of Tycho's supernova. 

The observed CO gas kinematics suggests that the large cavity is expanding at a velocity of $\sim$\,3-4\,km\,s$^{-1}$. 
The estimated velocity ($\sim$\,800\,\kms, with a mass-loss rate of $\sim$\,10$^{-6}$\,$M_{\odot}$\,yr$^{-1}$) and timescale 
($\sim$\,4-5\,$\times$\,10$^6$\,yr) of the wind needed for creating such a cavity are consistent with the predictions from 
the wind-regulated accretion model. 

Nevertheless, we note that the possibility of the random distribution of a large cavity around Tycho's SNR cannot be ruled 
out thus far. Further observations are needed to confirm the physical association of the large cavity with the remnant, as well 
as to comprehensively understand the nature of the streams and inner rim found therein. If the large cavity is really associated 
with the remnant, our result may imply that Tycho's supernova, the prototypical Type~Ia supernova in the Milky Way, arose from 
accretion onto a white dwarf.

\section*{ACKNOWLEDGMENTS}

We thank the anonymous referee for providing insightful suggestions and comments, which helped us to improve this work.
We are grateful to all the members of the Milky Way Imaging Scroll Painting CO line survey group, especially the staff of Qinghai
Radio Station of PMO at Delingha for support during the observations. This work was supported by the National Natural Science 
Foundation of China (grants Nos. 11473069, 11233007, and U1431231) and the Strategic Priority Research Program of the Chinese 
Academy of Sciences (grant No. XDB09000000). X.C. acknowledges the support of the Thousand Young Talents Program of China.

\clearpage
\begin{table}[ht]
\begin{center}
{Table~1. The H$_{2}$ gas properties of the objects around Tycho's SNR}
\begin{tabular}{l c c c c c} 
\hline\hline 
Object & Column density$^a$ & Gas mass$^a$ & & Column density$^b$ & Gas mass$^b$\\
 & ($\times$\,10$^{20}$\,cm$^{-2}$) & ($\times$\,10$^{3}$\,$M_{\rm Sun}$)  & &  ($\times$\,10$^{20}$\,cm$^{-2}$) & ($\times$\,10$^{3}$\,$M_{\rm Sun}$) \\ [0.2ex] 
\hline 
SE cloud                    & 11.9$\pm$2.4    & 21.9$\pm$4.4  & &     6.1$\pm$1.2    & 3.0$\pm$0.6   \\ 
NE cloud                    & 10.7$\pm$2.1    & 6.9$\pm$1.4   & &      7.7$\pm$1.5   & 1.0$\pm$0.2    \\ 
West cloud                 & 8.6$\pm$1.7     &  9.3$\pm$1.9  & &       6.6$\pm$1.3        & 0.5$\pm$0.1    \\ 
Stream$_{\rm SE}$    & 7.9$\pm$1.6     & 0.49$\pm$0.10 & &    3.9$\pm$0.8        & 0.05$\pm$0.01   \\ 
Stream$_{\rm NE}$    & 7.8$\pm$1.6     & 0.56$\pm$0.11 & &    4.7$\pm$0.9      & 0.05$\pm$0.01   \\ 
Stream$_{\rm NW}$   & 5.3$\pm$1.1     & 0.48$\pm$0.10 & &    2.9$\pm$0.6        & 0.01$\pm$0.005   \\ 
Inner rim                    & 8.9$\pm$1.8     & 1.3$\pm$0.3    & & 4.3$\pm$0.9        &  0.1$\pm$0.02 \\ 
Cavity                         & 1.5$\pm$0.3     & 2.7$\pm$0.5    & & 0.9$\pm$0.2       &  2.0$\pm$0.4  \\
\hline
\end{tabular}\\
$^a$Results estimated from the MWISP $^{12}$CO observations and CO-to-H$_2$ factor 
(1.8\,$\times$\,10$^{20}$\,cm$^{-2}$\,K$^{-1}$\,km$^{-1}$\,s; see text).\\
$^b$Results estimated from the MWISP $^{12}$CO and $^{13}$CO observations, assuming LTE condition.
\end{center}
\end{table} 


\clearpage

\begin{figure*}[!htbp]
\centering
\includegraphics[width = 0.95\textwidth]{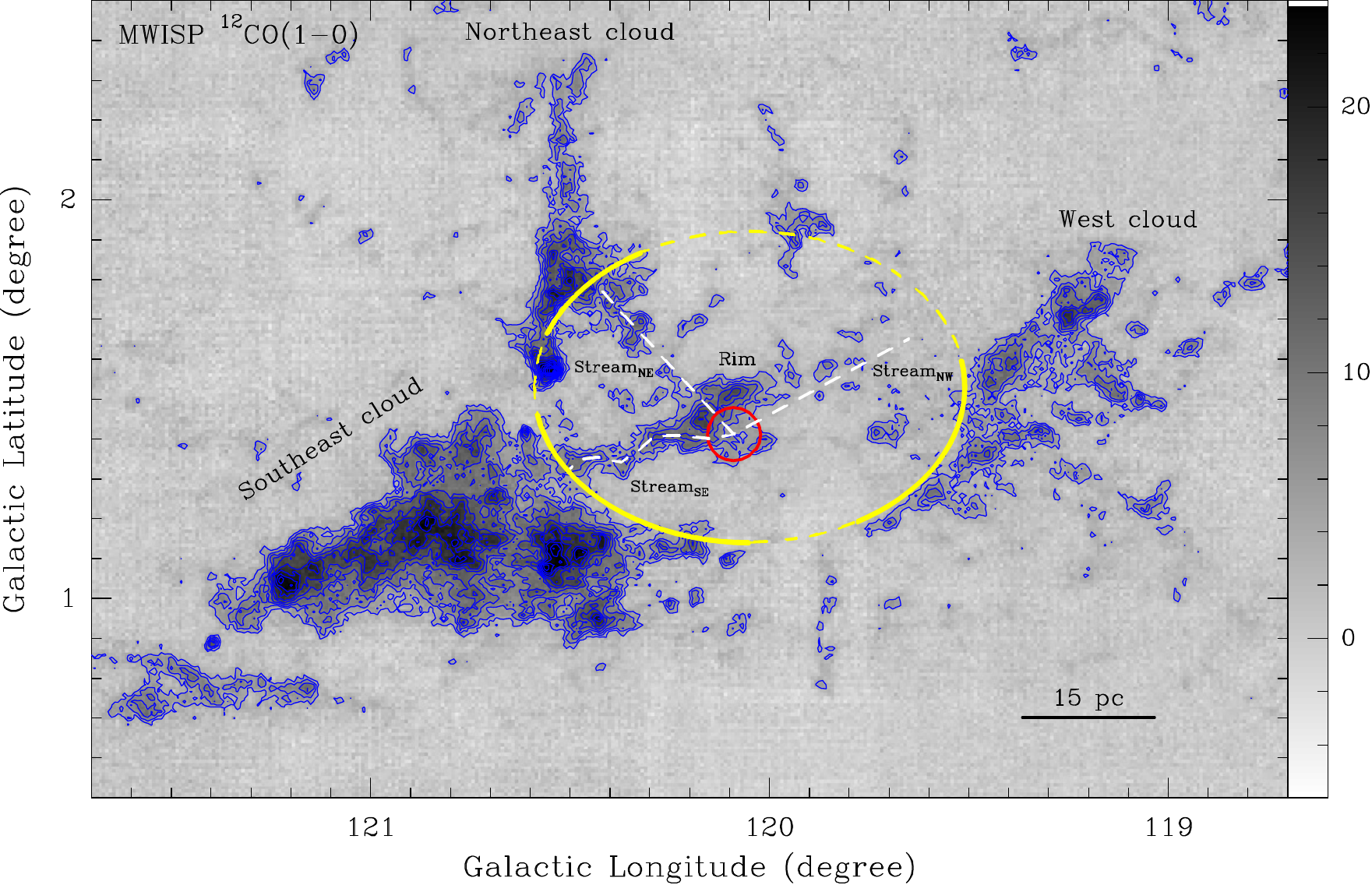}
\caption{The MWISP $^{12}$CO\,(1--0) velocity-integrated intensity image of Tycho's SNR. The unit of the scale bar is K\,km\,s$^{-1}$.
The emission is integrated between $-$68 and $-$57\,km\,s$^{-1}$, and the contours start at 5\,$\sigma$ and then increase in steps of
3\,$\sigma$ (1\,$\sigma$\,$\sim$\,0.8\,K\,km\,s$^{-1}$). The yellow solid lines show the arc-like structures observed in the three surrounding 
clouds around the remnant. The three arcs could be connected by the dashed lines and fitted into one complete ellipse, outlining a large 
cavity around the remnant. The three white dashed lines are shown to guide the eye for the stream-like structures seen in the cavity. The 
red circle shows the size (8$'$ in diameter) and position of the shell-like structure produced by the shockwaves from the supernova explosion, 
which was detected in the previous X-ray observations toward Tycho's SNR (e.g., Warren et al. 2005).\label{MWISP_12CO}}
\end{figure*}

\begin{figure*}
\begin{center}
\includegraphics[width=0.95\textwidth]{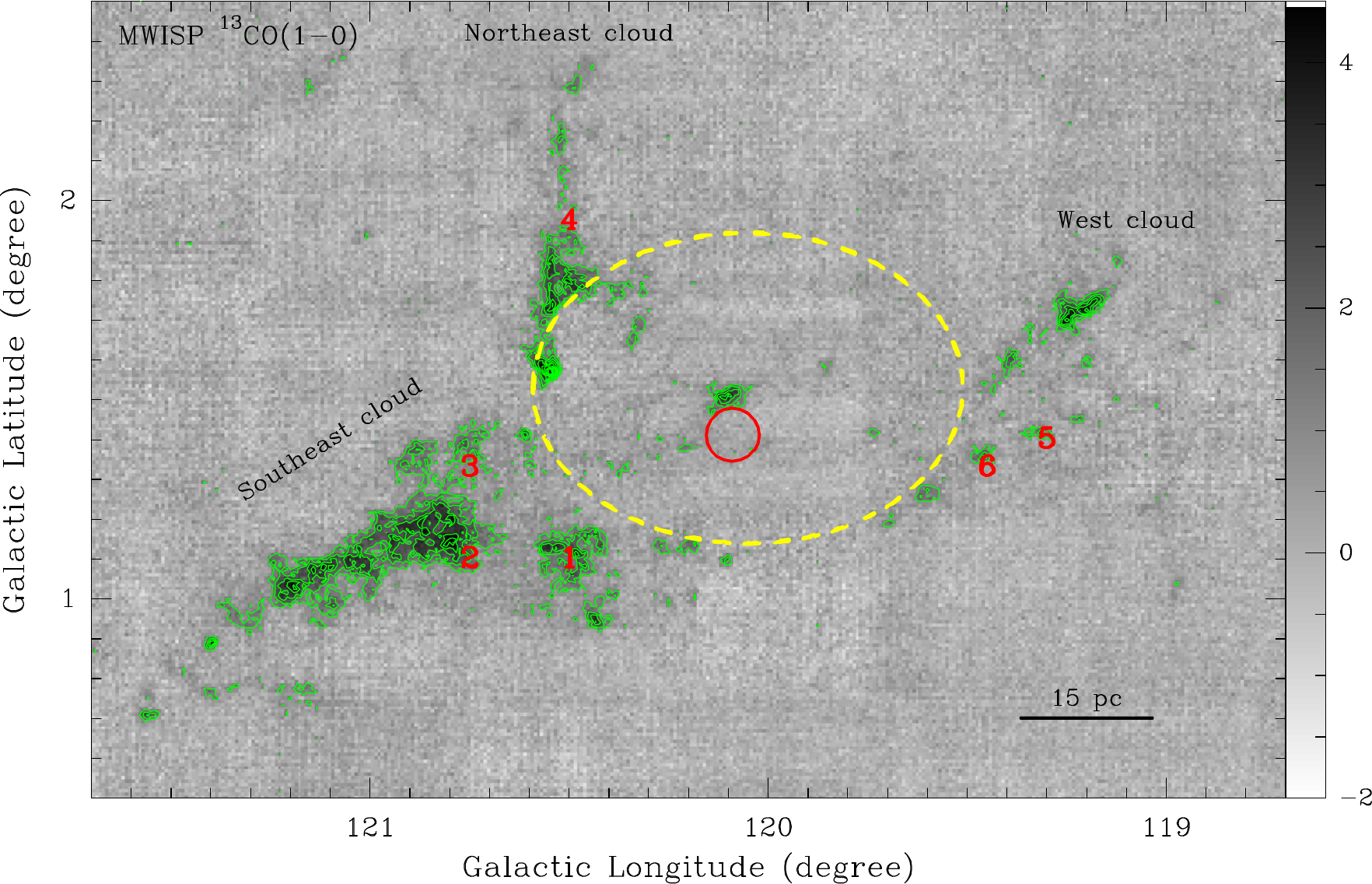}
\caption{The MWISP $^{13}$CO\,(1--0) velocity-integrated intensity image of Tycho's SNR. The unit of the scale bar is K\,km\,s$^{-1}$.
The emission is integrated between $-$65 and $-$59\,km\,s$^{-1}$, and the contours start at 3\,$\sigma$ and then increase in steps of
2\,$\sigma$ (1\,$\sigma$\,$\sim$\,0.4\,K\,km\,s$^{-1}$). The red circle shows the position and size of the shell-like structure in Tycho's 
SNR, while the yellow dashed ellipse shows the cavity found in the MWISP $^{12}$CO images. The numbers in the image mark the positions 
for sampling the CO spectra of the clouds.\label{MWISP_13CO}}
\end{center}
\end{figure*}

\begin{figure*}
\begin{center}
\includegraphics[width=0.95\textwidth]{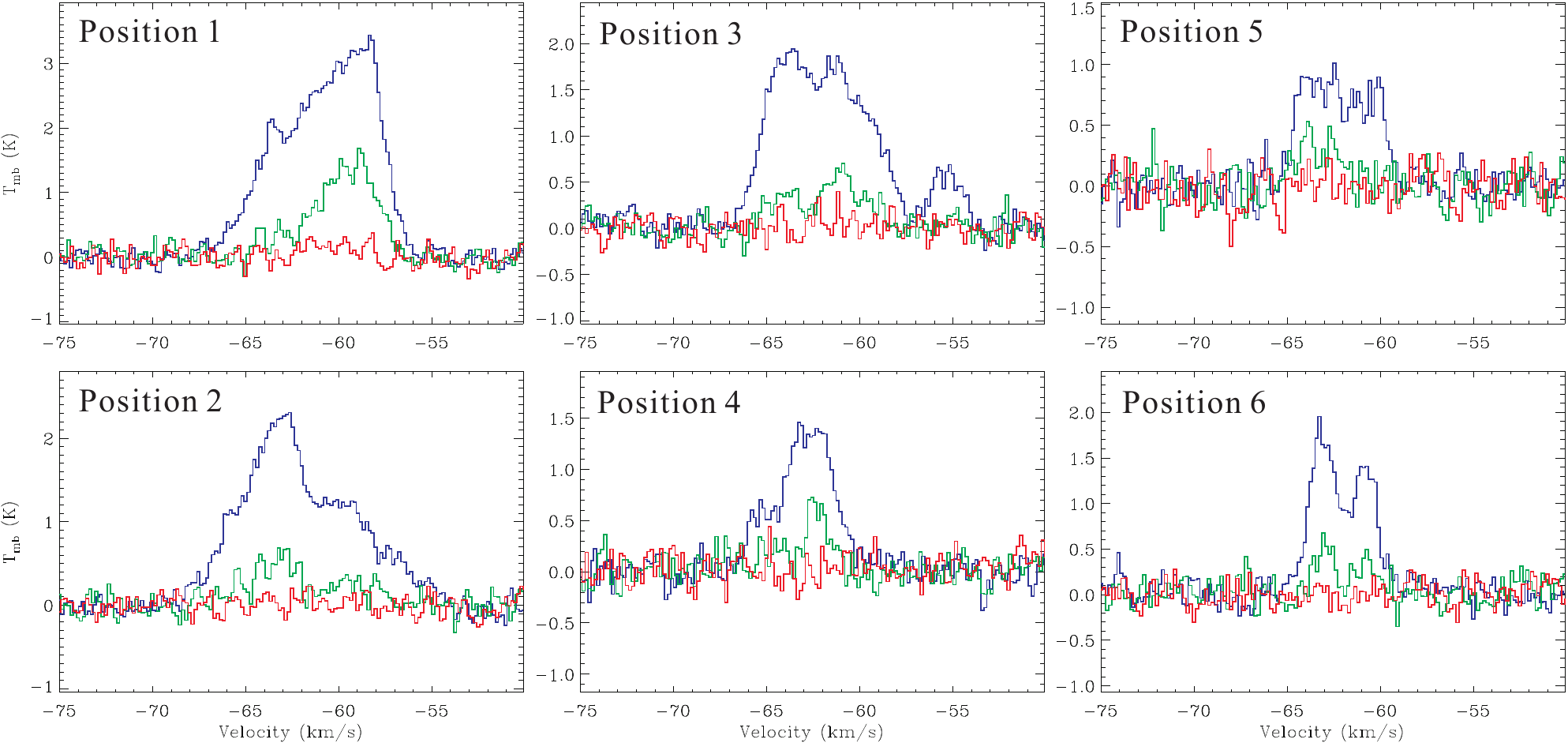}
\caption{The MWISP CO spectra in the surrounding clouds of Tycho's SNR. The spectra are sampled from the six positions marked in 
Figure~2, and averaged over an area of 0.1$^\circ$\,$\times$\,0.1$^\circ$. In each panel, the blue, green and red spectra represent the 
emission from the $^{12}$CO, $^{13}$CO, and C$^{18}$O lines, respectively.\label{MWISP_spectra}}
\end{center}
\end{figure*}

\begin{figure*}
\begin{center}
\includegraphics[width=0.95\textwidth]{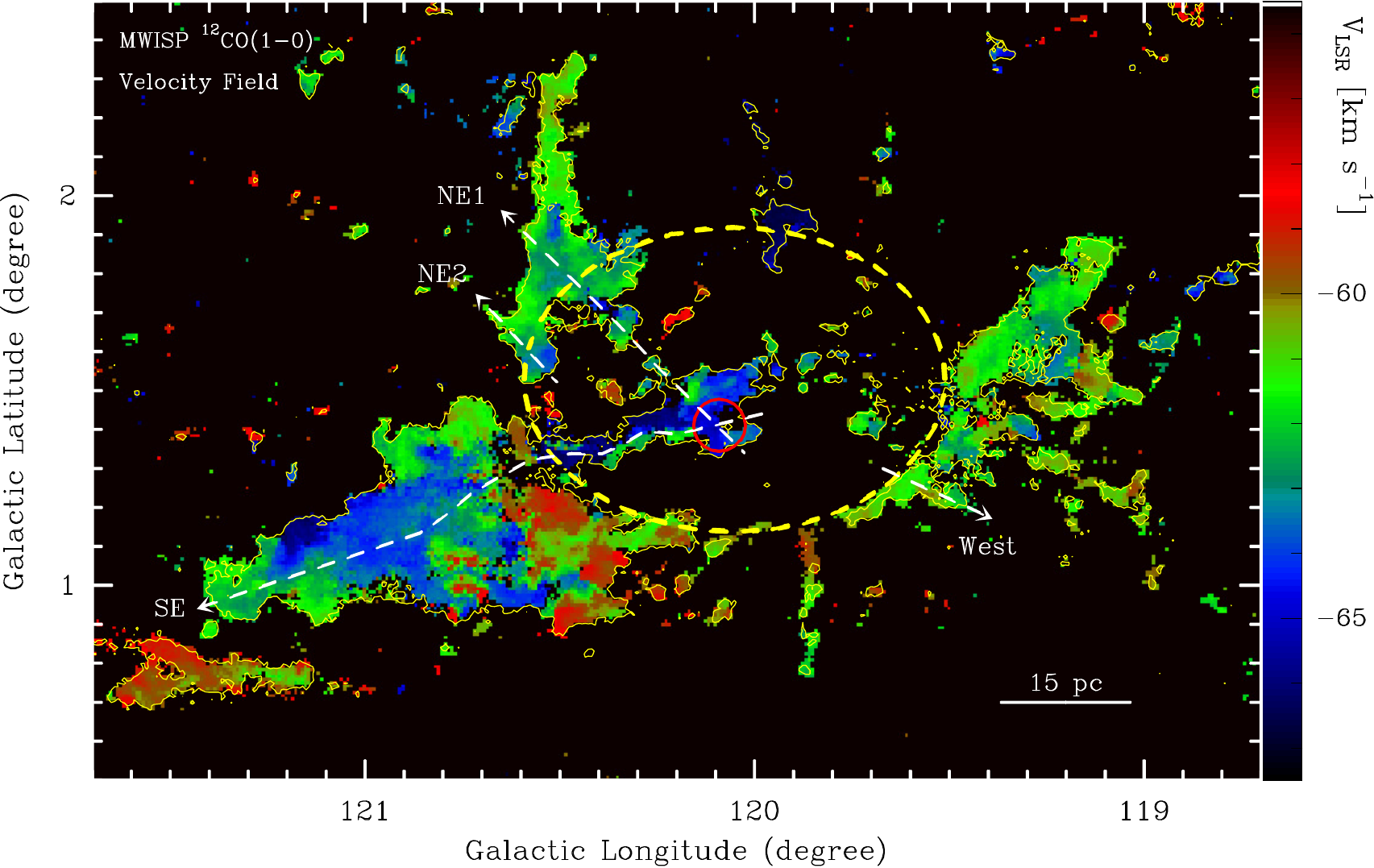}
\caption{The velocity field of the MWISP $^{12}$CO\,(1--0) emission (1st moment; color scale). The emission is integrated between $-$68 
and $-$57\,km\,s$^{-1}$. The yellow contour shows the intensity emission at the 5\,$\sigma$ level (1\,$\sigma$ $\sim$\,0.8\,K\,km\,s$^{-1}$). 
The yellow dashed ellipse shows the cavity found in the $^{12}$CO image, while the red circle shows the position and size of the shell-like 
structure found in Tycho's SNR. The four white dashed arrow lines show the routes of the position-velocity diagrams shown in Figure~5.
\label{velocity_field}}
\end{center}
\end{figure*}

\begin{figure*}
\begin{center}
\includegraphics[width=0.95\textwidth]{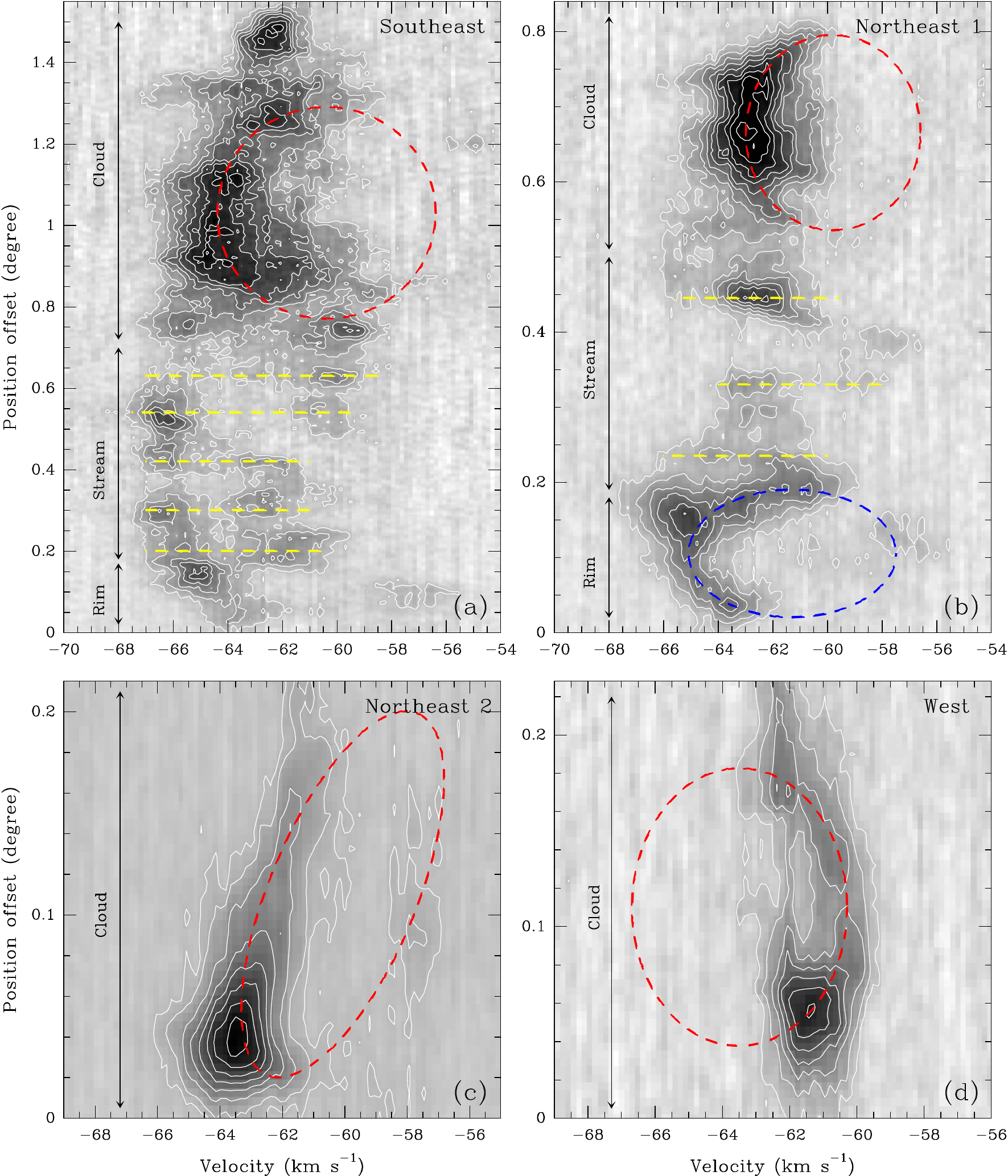}
\caption{Position-velocity diagrams along the four routes shown in Figure 4. (a) The contours start at 0.6\,K and increase in steps of 0.5\,K 
(1$\sigma$\,$\sim$\,0.15-0.16\,K). The red ellipse (with a velocity radius of $\sim$\,4\,\kms) shows a fitting toward the curve-shaped morphology 
seen in the southeast cloud (suggesting gas expansion), while the yellow lines show the velocity broadenings seen in the stream-like structure 
(suggesting shocked emission). (b) The contours start at 0.5\,K and increase in steps of 0.5\,K (1$\sigma$\,$\sim$\,0.15\,K). The red ellipse 
(velocity radius of $\sim$\,3.2\,\kms) and blue ellipse (velocity radius of $\sim$\,3.8\,\kms) show the fitting toward the curve-shaped morphologies 
seen in the northeast cloud and inner rim, respectively. (c) The contours start at 0.5\,K and 1.0\,K, and then increase in steps of 1.0\,K 
(1$\sigma$\,$\sim$\,0.15\,K). The velocity radius of the red ellipse is $\sim$\,3.3\,\kms. (d) The contours start at 0.5\,K and increase in steps 
of 0.5\,K (1$\sigma$\,$\sim$\,0.16\,K). The velocity radius of the red ellipse is $\sim$\,3.2\,\kms.
\label{PV1}}
\end{center}
\end{figure*}

\begin{figure*}
\begin{center}
\includegraphics[width=0.95\textwidth]{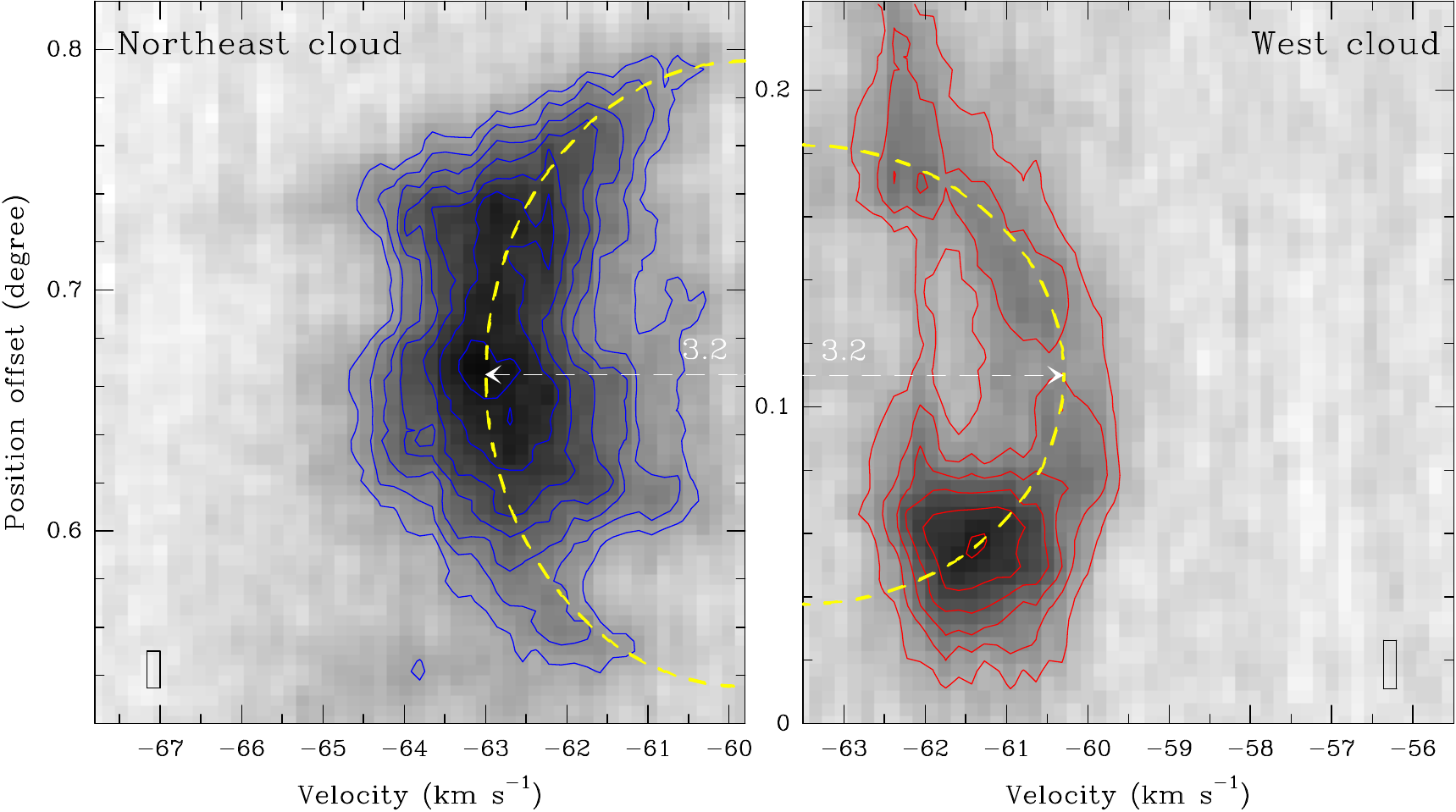}
\caption{A comparison between the position-velocity diagrams of the northeast cloud (from Fig.\,5b, but contours start at 1.5\,K and then 
increase in steps of 0.5\,K, where 1$\sigma$ is $\sim$\,0.15\,K) and the west cloud (from Fig.\,5d, but contours start at 1.0\,K and then 
increase in steps of 0.5\,K, where 1$\sigma$ is $\sim$\,0.16\,K). Both diagrams could be fitted with an ellipse (yellow dashed curve) with a 
velocity radius of $\sim$\,3.2\,\kms. The small rectangle in each panel shows the angular and velocity resolutions in the MWISP $^{12}$CO 
observations.
\label{PV2}}
\end{center}
\end{figure*}

\begin{figure*}[!htbp]
\begin{center}
\includegraphics[width=0.95\textwidth]{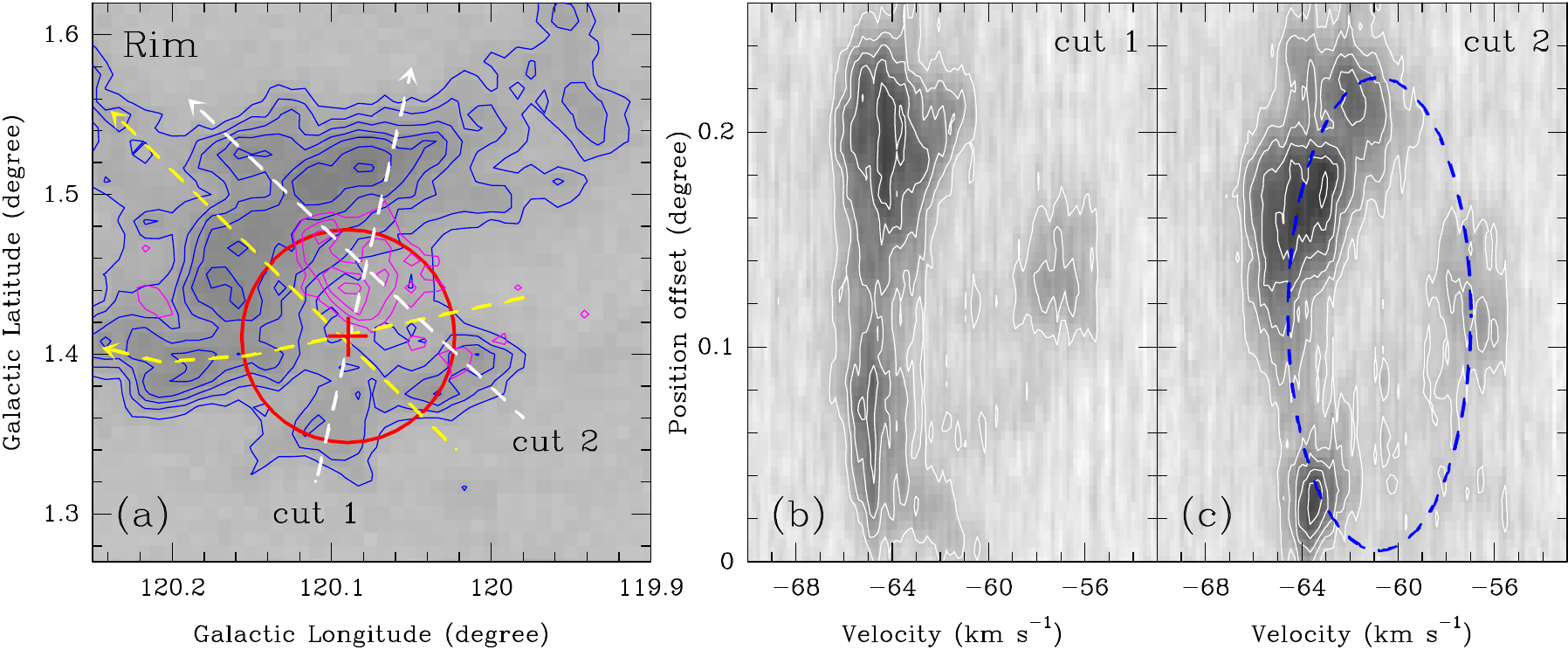}
\caption{(a) The MWISP $^{12}$CO intensity map of the inner rim. The blue and pink contours show the emission integrated between [$-$66, $-$60] 
and [$-$58, $-$55]\,\kms, respectively. The blue contours start at 5\,$\sigma$ and increase in steps of 3\,$\sigma$ (1\,$\sigma$\,$\sim$\,0.6\,K\,\kms), 
while the pink contours start at 3\,$\sigma$ and increase in steps of 2\,$\sigma$ (1\,$\sigma$\,$\sim$\,0.47\,K\,\kms). The red circle shows the position
and size of the shell-like structure (8$'$ in diameter). The yellow dashed arrow lines show the cuts toward the southeast and northeast clouds (see 
Figs.\,4 \& 5). (b-c) Position-velocity diagrams along the two cuts marked in panel `a' (white dashed arrow lines). The contours in the two diagrams start 
at 0.5\,K and increase in steps of 0.5\,K (1$\sigma$\,$\sim$\,0.14\,K). The blue ellipse in panel `c' shows a fitting toward the diagram with a velocity 
radius of $\sim$\,3.8\,\kms.\label{rim}}
\end{center}
\end{figure*}

\clearpage

\begin{appendix}

\section{MWISP CO\,(1--0) velocity channel maps}

Fig.\,A.\,1 shows the velocity channel maps of the MWISP $^{12}$CO emission. In the channel maps, the integrated velocity range 
is written in the top right corner of each panel (in km\,s$^{-1}$). The red circle shows the position and size of the shell-like structure 
in Tycho's SNR (8$'$ in diameter), while the yellow dashed ellipse shows the cavity found in the MWISP CO images.

\begin{figure*}
\begin{center}
\includegraphics[width=0.95\textwidth]{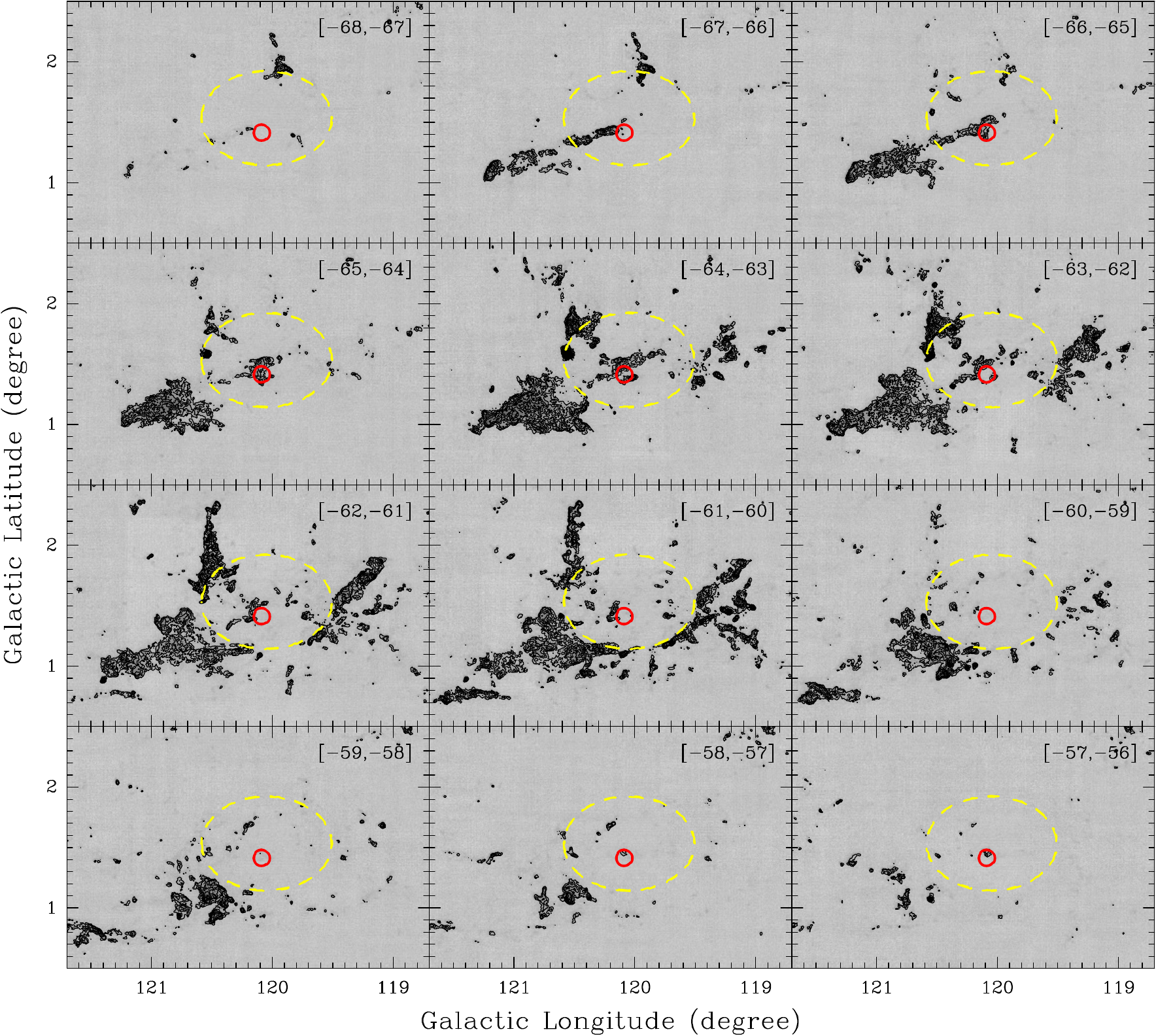}
\caption{The velocity-integrated intensity channel maps of the MWISP $^{12}$CO\,(1--0) emission. In each panel, contour levels start 
at 5\,$\sigma$, then increase in steps of 3\,$\sigma$, where the 1\,$\sigma$ level is $\sim$\,0.25\,K\,km\,s$^{-1}$. The integrated velocity 
range is written in the top right corner of each panel (in km\,s$^{-1}$). The red circle shows the position and size of the shell-like structure 
in Tycho's SNR, while the yellow dashed ellipse shows the cavity found in the MWISP CO images.\label{MWISP_channel_12CO}}
\end{center}
\end{figure*}

\section{Comparison with previous CO survey}

Fig.\,B.\,1 shows a large-field FCRAO $^{12}$CO\,(1--0) intensity image around Tycho's SNR, integrated with the same 
velocity range as the MWISP $^{12}$CO image shown in Fig.\,1. In the FCRAO survey, the three large clouds are also detected 
around the remnant, although the arc-like structures in the clouds are not as clear as those seen in the MWISP CO images 
(the 1\,$\sigma$ noise is $\sim$\,1.7\,K\,km\,s$^{-1}$ in the FCRAO survey, about 2 times larger than that in the MWISP 
survey). 

In the FCRAO CO image, we try to search for `cavities', suggested by arc- or ring-like structures, with the visual identification 
(as we did in the MWISP CO images). In a field with dimensions of 18$^\circ$\,$\times$\,8.4$^\circ$, another eight `cavities' 
are found (see Fig.\,B.\,1), though none of these cavities are stronger in intensity (the intensities summed from the surrounding 
arc- or ring-like structures) than the one found around Tycho's SNR. If we assume any cavity closer than 0.6$^\circ$ (major 
radius of the cavity found in the MWISP CO images) to the remnant as an association, the probability for a cavity to be detected 
by chance is approximately [8/(18$\times$8.4)]\,$\times$\,0.6$^2$$\pi$\,$\approx$\,0.06. The probability would be higher, 
if taking broader velocity ranges into account.

\begin{figure*}[!htbp]
\begin{center}
\includegraphics[width=0.95\textwidth]{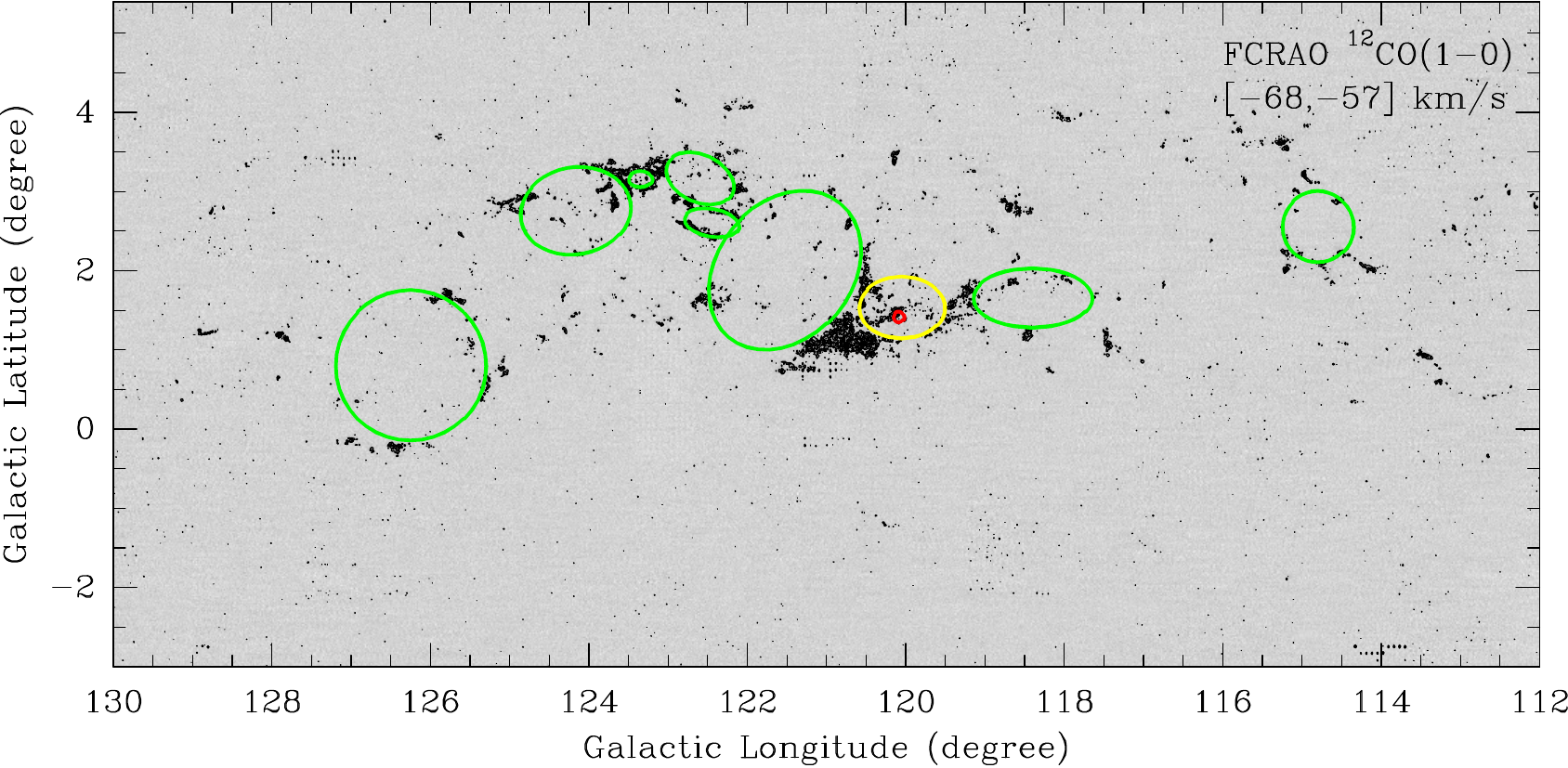}
\caption{The larger-field FCRAO $^{12}$CO\,(1--0) intensity image around Tycho's SNR, integrated between $-$68 and $-$57\,km\,s$^{-1}$ 
(as same as Fig.\,1). The contours start at 3\,$\sigma$ and then increase in steps of 2\,$\sigma$ (1\,$\sigma$\,$\sim$\,1.7\,K\,km\,s$^{-1}$). The 
red circle shows the shell-like structure in Tycho's SNR, while the yellow ellipse shows the cavity found in the MWISP CO images. The green 
ellipses show similar cavity-like structures found in the FCRAO CO image.{}\label{FCRAO}}
\end{center}
\end{figure*}

\end{appendix}

\end{document}